\documentclass[10pt]{blois07}
\usepackage{graphicx}
\usepackage{cite,./mcite}

\begin{document}
\title{The soft and the hard pomerons: elastic scattering and unitarisation}
\author{J.R. Cudell$^1$\protect\footnote{\ \  talk presented at EDS07},
A. Lengyel$^2$, E. Martynov$^3$, O.V. Selyugin$^4$}
\institute{$^1$JR.Cudell@ulg.ac.be, AGO dept., Universit\'e de Li\`ege, 4000 Li\`ege, Belgium\\ 
$^2$sasha@len.uzhgorod.ua, Inst. of Electron Physics,  88000
Uzhgorod, Ukraine,\\ 
$^3$martynov@bitp.kiev.ua,  Bogolyubov Institute for Theoretical Physics, 03143 Kiev, Ukraine,\\
$^4$selugin@theor.jinr.ru, Bogoliubov Theoretical Laboratory,  141980 Dubna,  Russia}
\maketitle
\begin{abstract}
The hard pomeron component needed to reproduce small-$x$ data seems to be present 
in elastic scattering at moderate energy. If this is the case, it is likely that 
the total cross section at the LHC will be appreciably larger than previously 
expected.
\end{abstract}

\section{The effective trajectory}
Historically, hadronic exchanges have been remarkably well described by the 
simple-pole model of Donnachie and Landshoff \cite{DLsoft}, which assumes that, at high energy, 
the hadronic amplitude is dominated by a simple-pole pomeron:
\begin{equation}
A(s,t)=C(t)s^{1.08+0.25t}
\end{equation}
The first question is whether this simple model extends to other processes, and in 
particular to those measured at HERA.

H1 and ZEUS have done amazingly good and useful measurements, and extracted very 
precise information on $F_2$, as well as on pomeron-dominated processes such as 
quasi-elastic vector-meson production, or DVCS. This means that we have a very 
precise picture of the pomeron, and of the dependence of its properties on $Q^2$ 
and $t$. If we represent pomeron exchange by a simple pole corresponding to a 
trajectory $\alpha(t)=\alpha(0)+\alpha' t$, its contribution to $F_2$ would be proportional to 
$x^{\alpha(0)-1}$ while its contribution to quasi-elastic scattering would go as 
$\sigma\sim W^{4(\alpha(t)-1)}$. All the data lead to the following effective properties (see e.g. \cite{hpdat1,*hpdat2}):
\begin{itemize}
\item the intercept varies with $Q^2$ and with the mass $M$ of the produced meson. Its 
value is always constrained by $1.08<\alpha(0)<1.45$, and increases with $Q^2$ and 
$M^2$.
\item The slope $\alpha'$ of the trajectory is not universal: it is lower in processes 
containing a hard scale, such as J/$\psi$ production, and is constrained by 
$0.16$ GeV$^{-2}>\alpha'>0.08$ GeV$^{-2}$.
\end{itemize}
Unfortunately, because the masses and $Q^2$ are external variables, it is not 
possible that the trajectory associated with a simple pole depends on them. One 
may get such a behaviour via more complicated singularities in the complex $j$ 
plane. Nevertheless, it is striking to observe that different processes lead to 
the same bounds on the effective intercept, and that both intercept and slope join 
smoothly with the original Donnachie-Landshoff soft pomeron. 
\section{Regge theory and the two-pomeron picture}
This observation prompted Donnachie and Landshoff to extend their original model 
by adding a new contribution to it \cite{DLhard}: the hard pomeron.
The hadronic elastic amplitude thus becomes
\[
A_{2\rightarrow2}(s,t)=\sum_{i}C_{i}(Q^{2},M^{2},t)s^{\alpha_{i}(t)}\]
In general, several trajectories are needed to reproduce all the data, but the two 
dominant ones at high energy are the hard and the soft pomerons. Both are supposed 
to have universal trajectories. The effective intercept thus comes from mixing 
these two pomerons with coefficients that depend on $Q^2$ and $M$. Following these 
ideas, Donnachie and Landshoff were able to reproduce all the data from HERA. The 
hard pomeron intercept was fitted to $\alpha_H(0)=1.4372$, so that the small-$x$  
measurements of $F_2$ were reproduced \cite{DLsmallx}. The fit was then extended to higher values 
of $x$ \cite{DLx}, and was made compatible with DGLAP evolution \cite{DLevol}, so that partonic 
distributions could be obtained from it \cite{DLglu}. These ideas were also 
applied to vector-meson quasielastic photoproduction \cite{DLmes}, 
which are also well reproduced in a 2-pomeron 
model, when the couplings of the pomerons depend on the mass of the produced 
meson. The fits to these processes, which now involve an interference between the 
hard and the soft pomeron, lead to compatible values for the hard-pomeron 
intercept, and to an estimate for its slope: $\alpha'_H\approx 0.1$ GeV$^{-2}$, 
while the soft pomeron trajectory remains as in Eq. (1).

So the addition of a hard pomeron leads to a good description of the data at HERA. 
But if the hard pomeron is a universal exchange, then one would expect it to be 
present even in soft data. It is indeed unlikely that the couplings become exactly 
zero for all on-shell light hadrons, especially if we remember that they vary in 
photoproduction with the mass of the produced mesons. This prompted us to 
reconsider the fits to total cross sections and to the real part of the elastic 
amplitude.

\section{The hard component of soft data}
\label{sec:xxx}
The main motivation of \cite{CLMS} was in fact to understand why the soft pomeron 
model of Eq. (1) could provide a good fit to total cross sections for 
$\sqrt{s}\geq 5$ GeV, while it failed to reproduce 
the data for $\rho(0)$ \cite{benchmarks}. Hence 
the paper used integral dispersion relations in a rather careful way, and tried to 
check whether lifting totally the degeneracy of the dominant meson trajectories 
would help. It was then found that if one added a new crossing-even trajectory, 
its intercept would automatically go to a value around 1.4, and the quality of the
fit would become comparable to that of the best parameterisations of 
\cite{benchmarks}. A careful analysis of the results shows that the main 
improvement comes from the fit to $\pi p$ and $K p$ data, while the $pp$ 
and $\bar p p$ fit is not affected. In fact the coefficient of the hard pomeron
is essentially zero in this case, which explains why this contribution was
not considered before \cite{medium}. The reason for this decoupling is clear:
the inclusion of a hard pomeron makes the cross sections rise very fast, so that
unitarity corrections must be considered between the ISR and the Sp$\bar{\rm p}$S
energies\footnote{See O.V. Selyugin's contribution to these proceedings.}. 

So our strategy was to fit data below 100 GeV, which gave us the following 
results:
\begin{itemize}
\item the hard pomeron intercept is $$\alpha_H=1.45\pm 0.01;$$
\item if we write the coefficients of the amplitude for $ab\rightarrow ab$ elastic 
scattering as $g_a g_b$, we obtain an inverse hierarchy: $$g_K\approx 1.1
 g_\pi\approx 3.2 g_p.$$ The hard pomeron seems to couple more to smaller objects.
\item As the amplitude is described by simple-pole singularities, its
terms should obey Regge factorisation, which can be tested by deriving the
amplitude for $\gamma\gamma$ scattering from the amplitudes for $pp$
and $\gamma p$ scattering. The resulting cross section does reproduce the
LEP measurement, favouring the deconvolution with PHOJET. The hard pomeron
singularity must then be rather close to a simple pole.
\end{itemize}

After this study, three of us considered the off-forward case \cite{CLM}. 
We limited ourselves to the first cone $|t|<0.5$ GeV$^2$, so that we would not 
have to deal with multiple exchanges, or with the question of the odderon, and 
found that many data sets had problems at small $|t|$, so that we also used $|t|
>0.1$ GeV$^2$. The elastic $ap$ amplitude is then given by
$$A_{2\rightarrow2}^{ap}=C_{H}^{ap}F_{H}^{p}(t)F_{H}^{a}
(t)s^{\alpha_{H}}+C_{S}^{ap}F_{S}^{p}(t)F_{S}^{a}(t)+{\rm meson}\:{\rm 
contributions},$$
with $F_i(0)=1$. 
The dataset we fitted to includes all the available data for $pp$, $\bar pp$, $Kp$ 
and $\pi p$ elastic scattering, which we extracted from the HEPDATA system 
\cite{HEPDATA}, checked, and compiled in a unique format. The resulting 
database\footnote{available at http://www.theo.phys.ulg.ac.be/\~cudell/data/}, 
contains about 10000 points, more than 2000 of which fall in the $t$ interval considered here, and at $\sqrt{s}\leq 100$ GeV. Because of this large number of points, we were able to extract the form 
factors $F_i(t)$ directly from the data, by fitting small intervals $t_j\pm \Delta t_j$ to constant $F_i(t_j)$, and by later fitting smooth curves to these values.
The resulting form factors are shown in Fig. 1.
\begin{figure}
\includegraphics{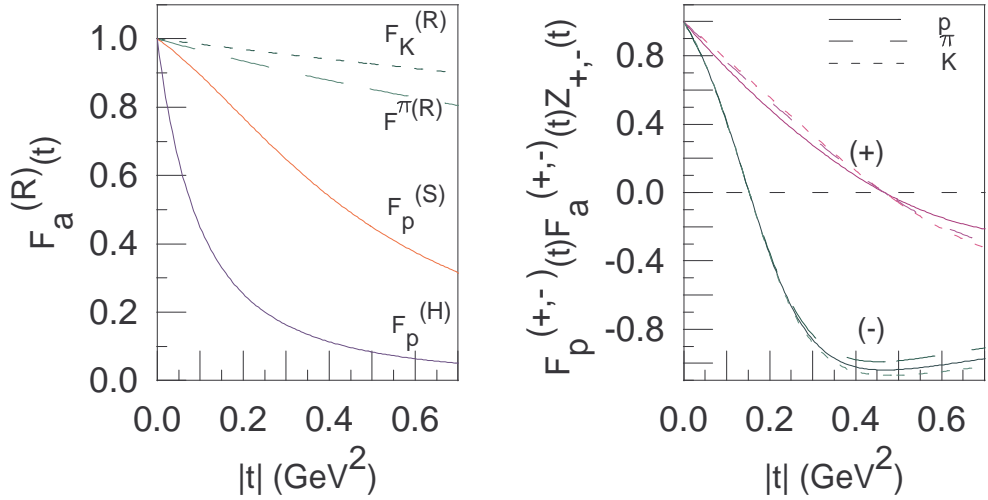}
\caption{Form factors for the various exchange terms, for $|t|\leq 0.5$ GeV$^2$.
The left figure gives the form factors of the two pomerons which are
different in the proton case, but taken as identical for mesons ($F^{(R)}$
is the pomeron form factor for both soft and hard pomerons in that case).
The right figure shows the form factor of the dominant 
$C=\pm 1$ meson exchanges.} 
\end{figure} 
There is enough data in the proton case to distinguish between the hard pomeron 
and the soft pomeron, but we had to assume they were degenerate in the $\pi p$ and 
$Kp$ cases. For the meson exchanges, we assumed that $C=+1$ and $C=-1$ trajectories were not degenerate, and found that both form factors 
had zeroes which we took into account by multiplying standard
form factors by a function $Z(t)$ which is asymptotically 1 but
changes sign at $t\approx -0.15$ GeV$^2$ for the $C=-1$ trajectory
and at $t\approx -0.47$ GeV$^2$ for the $C=+1$ meson tractory.

Using direct extraction of the form factors from data produces a remarkably good fit, with a $\chi^2/dof=0.95$, partially shown in Fig. 2. 
\begin{figure}
\includegraphics[width=0.8\textwidth]{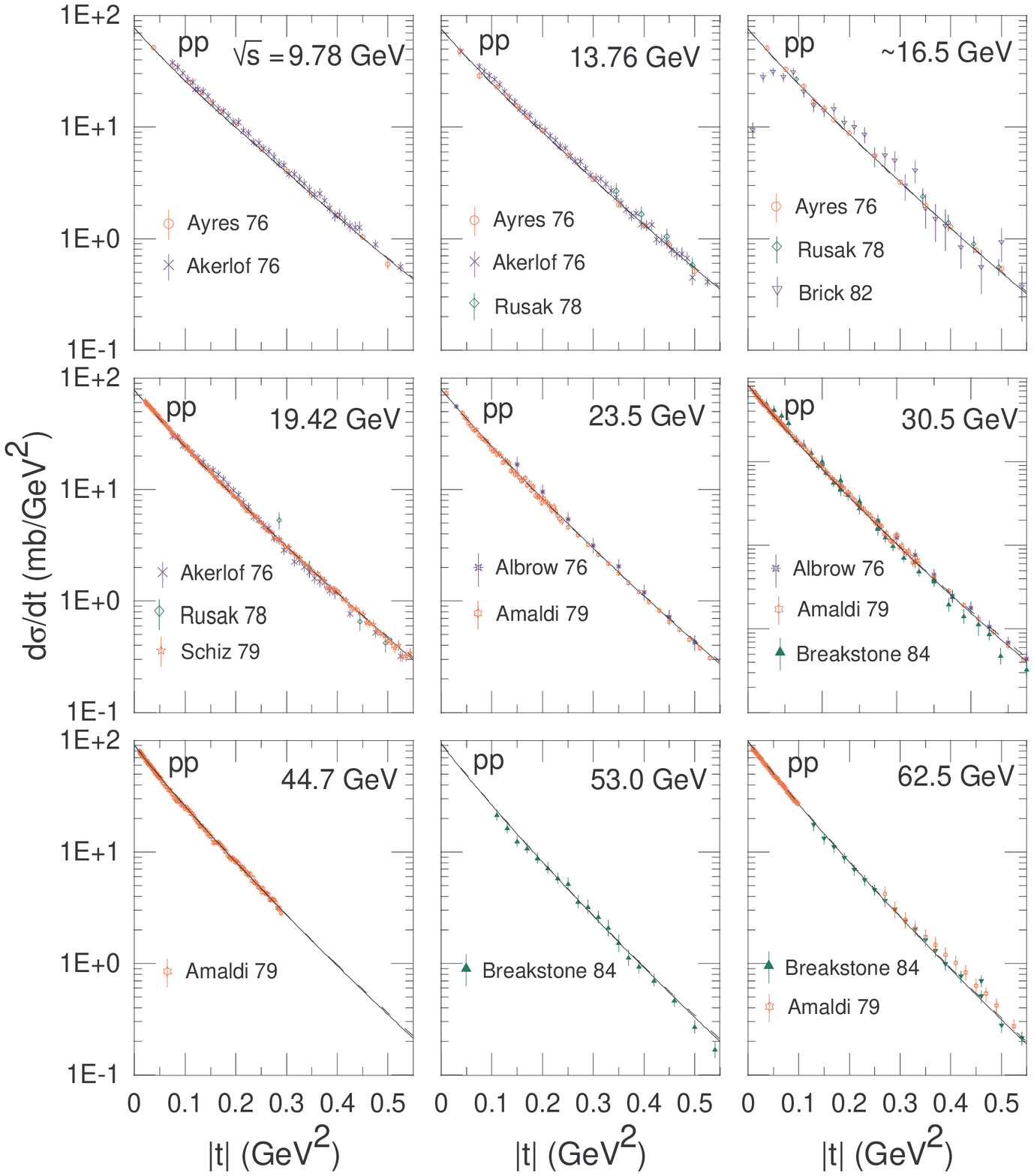}
\caption{Fit to elastic differential cross sections in the $pp$ case.} 
\end{figure}
Note that we had to exclude a few datasets which 
had gross disagreements with the others (about 100 points). Besides the form factors, it also produces trajectories for the pomerons. We find that
\begin{eqnarray}
\alpha_S&=&1.07+0.3 t\\
\alpha_H(t)&=&1.45+0.1 t
\end{eqnarray}
Although the central value of the hard pomeron slope is in perfect agreement with \cite{DLmes}, one must mention that it has a large error, as it is correlated with the choice of form factor for the hard pomeron.

\section{Unitarisation and high energies}
As explained above, we have had to consider the hard pomeron at low energy 
$\sqrt{s}\leq 100$ GeV. In order to make contact with higher-energy data, and with 
LHC physics, it is clear that multiple exchanges have to be considered. The
 problem of their inclusion is however far from solved.
Considerable progress has been made recently in addressing the unitarisation of the 
BFKL pomeron in a perturbative setting, based on the idea of the saturation of the 
gluon density. Although there seems to be some relation between the equations 
describing parton saturation and conventional unitarisation schemes \cite{CSunit}, 
it is far from clear whether one can extend these perturbative methods to a soft 
regime.

In \cite{benchmarks}, it was shown that a variety of analytic parameterisations 
leads to a total LHC cross section between 85 and 117 mb. We showed in \cite{CLMS} 
that a model including a hard pomeron could lead to a cross section compatible 
with these limits, provided one used an extended eikonal scheme. 

However, two of us have considered in \cite{CSpaper} more traditional schemes, 
i.e. the saturation of the profile function when it reaches the black-disk limit, 
or the standard one-channel eikonal. The former has the advantage to keep the 
simple-pole nature of the hard pomeron, so that the low-energy fit is not 
affected, whereas the latter implies a refitting, which however does not change 
the parameters significantly. Interestingly, both schemes predict a large cross 
section at the LHC, of the order of 150 mb, as show in Fig. 2.
\begin{figure}
\centerline{\includegraphics[width=0.5\textwidth]{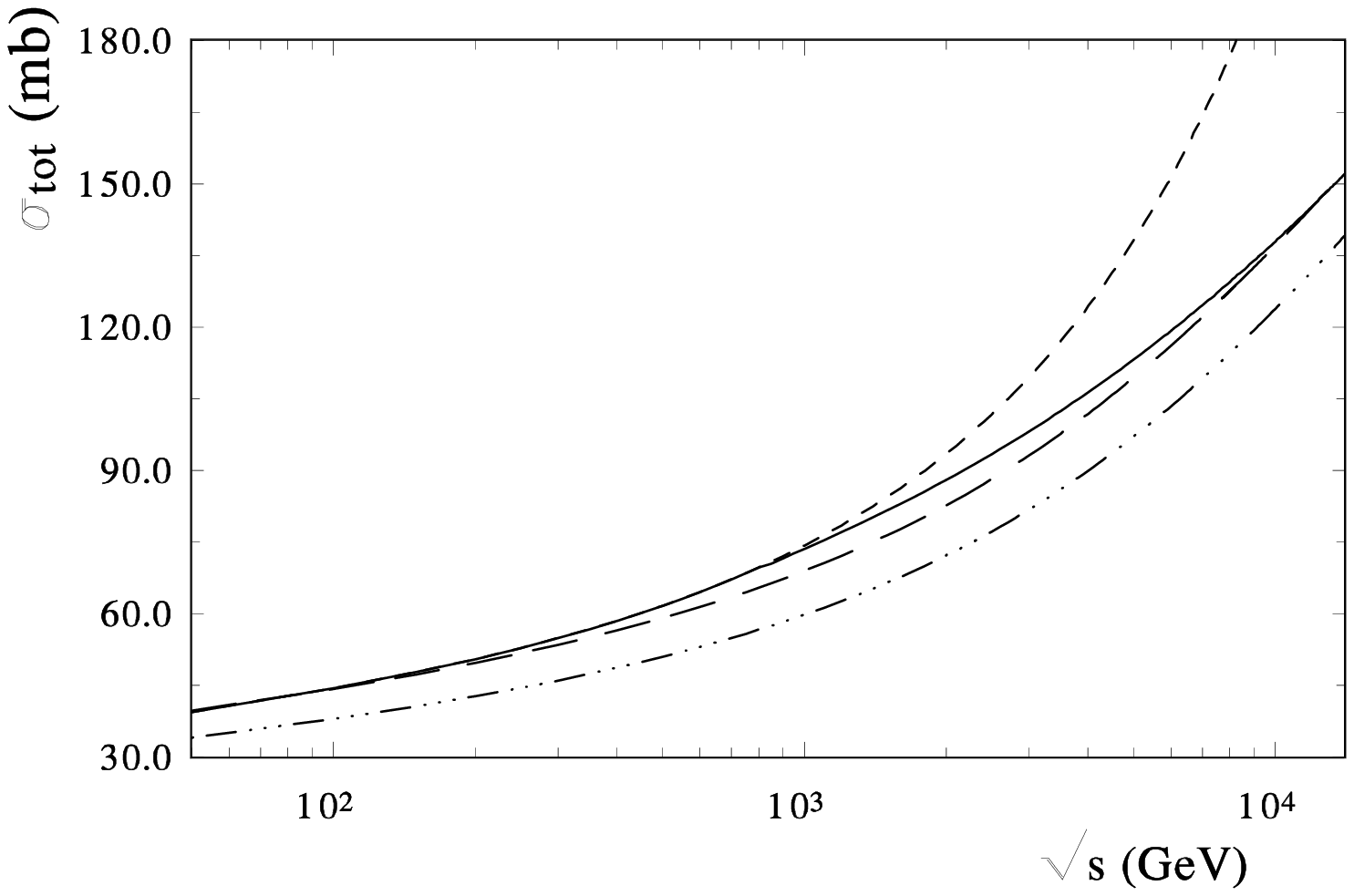}}
\caption{The total cross section as a function of $\sqrt{s}$, 
for the bare amplitude (short dashes), a saturated amplitude (plain curve),
an eikonalised amplitude (dash-dot-dot), and for a refitted eikonal 
(long dashes).} 
\end{figure} 
It thus seems possible to evade the bound presented in \cite{benchmarks}, provided 
a universal hard pomeron trajectory is present even in soft scattering. Note that 
such a trajectory would have consequences not only for the total cross section, 
but also for the value of the $\rho$ parameter and for the slope of the 
differential cross section $B(t)$, hence it is very important that all these
be measured at the LHC.
\section{Conclusion}
The idea of universal exchanges of Regge trajectories is an old one, which has 
been very fruitful in describing hadronic scattering data. It may in fact be
the only way to get a grasp on these processes, which escape perturbation theory.
The data of HERA have taught us that a hard singularity is present in 
$\gamma^{(*)}p$ scattering for large $Q^2$ and for large masses. It is natural
to assume that such a singularity is also present in other hadronic processes,
even at small $Q^2$ or for small masses. We have found that the inclusion of
a hard pomeron in fits to soft data  appreciably improves the description of the
data at energies below 100 GeV. The coupling of the hard pomeron is then only a 
few percents of that of the soft pomeron, but this is enough to reach the black 
disk limit around 1 TeV. One then needs to worry about multiple exchanges, and
this makes predictions for the LHC  quite uncertain.

One can say that the observation of a cross section lower than 117 mb would
disfavour the ideas presented here, as the unitarisation scheme needed would 
have to use strong rescatterings, which have no reason to disappear at low energy. 
On the other hand, the observation of a cross section higher than 117 mb would be 
a strong indication that a hard singularity is there in soft data. If this is the 
case, rescattering/saturation effects will modify not only $\sigma_{tot}$ but also 
$\rho$ and $B(t)$. It thus seems a good idea to devise a strategy to measure
these quantities in an independent way at the LHC.

\noindent{\small {\bf Acknowledgements:}  The authors would like to thank  
P.V. Landshoff for helpful discussions  and M. Diehl for his careful 
editing.
 O.S. gratefully acknowledges financial support
  from FRNS and would like to thank the  University of Li\`{e}ge
  where part of this work was done.
    }
\begin{footnotesize}
\bibliographystyle{blois07} 
{\raggedright
\bibliography{blois07-JRCudell}
}
\end{footnotesize}
\end{document}